\documentclass[12pt]{article}

\usepackage{graphicx}
\usepackage{amsfonts,amsmath,amssymb,amsthm}
\usepackage{indentfirst}
\usepackage{cite,hyperref}
\usepackage{mathptmx}
\usepackage{cleveref}

\numberwithin{equation}{section}
\DeclareMathAlphabet{\mathcal}{OMS}{cmsy}{b}{n}

\begin{document}

\title{M{\o}ller scattering in $2 + 1$ of generalized quantum electrodynamics in the heisenberg picture}
\author{David Montenegro$^{1}$\\
\textit{{$^{1}${\small Instituto de F\'{\i}sica Te\'orica (IFT), Universidade Estadual Paulista}}} \\
\textit{\small Rua Dr. Bento Teobaldo Ferraz 271, Bloco II Barra Funda, CEP
01140-070 S\~ao Paulo, SP, Brazil}\\
}
\maketitle
\date{}

\begin{abstract}
In this paper, we investigate from the framework of generalized electrodynamics the differential cross section of the electron-electron scattering process  $e^- e^- \rightarrow e^- e^-$, i.e., M{\o}ller scattering, in $(2+1)$ dimensions in the Heisenberg picture. To this goal, one starts within the stable and unitary framework of planar generalized electrodynamics, instead of Maxwell one. We argue the Haag's theorem strongly suggests the study of the differential cross section in the Heisenberg representation. Afterward, we explore the influence of Podolsky mass cutoff and calculate the differential cross section considering data based on condensed matter systems.
\end{abstract}

\newpage

\section{Introduction}

Quantum electrodynamics $(QED)$ is a gauge theory of remarkable success with highly attractive from the theoretical and experimental viewpoint. The satisfactory conciliation between them is possible due to the fact, among others, the perturbative evaluation of cross section in different levels of energy. The great accomplishment in almost everything we know concerning particles has been available because of scattering methods. Such a picture has been supported by phenomena in high energy as well as condensed matter that rely on different information through experiments. One of the most fundamental observations is the electron-electron interaction which in the perspective of M{\o}ller $(e^- e^-)$ scattering characterizes somewhat special since its large cross section contributes for a good statistic to examine the standard model and test new emergent physics \cite{holyshit}.

As it is well known, the $QED$ structure and symmetries are inherent restricted by the dimensionality of the system. Particularly, planar $QED$ has attracted attention as a potential framework to discuss a large class of systems in the perturbative regime. The recent advances suggest as a qualitative explanation for the mechanism generating the formation of cooper pair in high-$\textit{T}_c$ superconductors \cite{ferreira}. Other notable examples are the prominent features of the fractional quantum Hall effect \cite{Hall}, the Heisenberg model of quantum spin \cite{Dillenschneide}, the quantum simulator of cold atoms \cite{coldd}, and the band structure of graphene \cite{opletter}. Furthermore, we also consider as a toy model to explore dynamical chiral symmetry breaking in quantum chromodynamics (QCD) \cite{QCD}.

Despite the remarkable application in many areas, planar physics belongs to the group of effective field theory. Thus, at this point, we are in the position to generalize the quantum electrodynamics in $(2+1)$ dimensions $(QED_3)$ and include higher order terms in such a way to leave the original symmetries unchanged.

Higher order field equations (third derivative or higher at the level of dynamics) have been present in many physical situations, for example, the extension of gravity models \cite{gravityyy}, the coupling of super string theory  \cite{superstringtheory}, and supersymmetry (SUSY) \cite{susy}. It seems an attractive way to investigate questions where low energy theories are insufficient to archive, here low energy corresponds to a defined scale in the system. This scenario may then lead us to assume that these models are incomplete or inaccurate. Higher derivative field theories draw useful insights in the clarification of underlying fundamental aspects of low-energy models and there is a wide class of new higher order theories that we can extract information and obtain interesting consequences. One remarkable misleading should be highlighted in this procedure before mentioning the aim of our paper. A first approach could induce that the original theory would appear small perturbed with higher derivative corrections even if assuming a small coupling. However, in contrast to lower derivative corrections, it is possible to recognize along with this paper that the higher theories are entirely different from the original ones and more fundamental due to the increase of new degrees of freedom and families of solutions.


At early in 1940, Podolsky and Schwed proposed \cite{Podolskyorigin} an extension of Maxwell-Lorentz electrodynamics. Initially, they intended to lead with classical problems as the $4/3$ factor in Abrahaam-Lorentz theory \cite{43problem} and remove the infinity from the self-energy charge particle ($r^{-1}$ singularity) \cite{Podolskyorigin}. The Podolsky's non-singular Lagrangian exhibits a natural regulator (cutoff) encoded in the massive parameter, which becomes relevant from classical and quantum view and admits, as a natural extension of Maxwell electrodynamics $(ME)$, a strength field quadratic divergent term, which conserves the linearity (superposition principle), Lorentz, and Gauge invariance. We can ask if assuming these properties, the $GQED$ would be unique. One rigorous way to answer this question is by Utiyama’s systematic method that proved the uniqueness apart from a surface term \cite{Cuzinattogauge}.  The Podolsky theory is often called generalized quantum electrodynamics $(GQED)$.

The novel feature inherent from any Higher derivative Lagrangian is the unboundedness of the canonical Noether energy that yields dynamic instability and negative norm states at the classical and quantum level, respectively. Ostrogradski has already evidenced this undesirable behavior when developed a canonical program to deal with non-degenerate higher derivative Hamiltonian \cite{Ostrogradski}. However, as pointed in \cite{russo}, it should be stressed that the positivity of energy is not a necessary but a sufficient requirement for the stability of equations of motion. Following this line of thought, we should update the criteria of energy positivity from Noether's theorem to validate the stability. It has been demonstrated in \cite{russo} the formalism of the Lagrangian anchor established the proper connection between the positive integral of motion and time translation invariance, so the boundness motion ensures classical and quantum stability. Also, this enables us to identify a safe room of stable higher order derivative field theories even if their unboundedness Noether's energy restricts them, as the Podolsky one. Moreover, the BRST symmetry revealed Podolsky as a unitary model \cite{Nogueira}. The fact  $GQED$  also assures unitarity and stability turns out a suitable candidate to deal with fermion and photon interaction.

Scattering theories provide the bridge for comparison between experimental data and quantum field theory, however, a mathematical problem may concern the characterization of these techniques. It is not surprising that we can construct a unitary and invertible mapping of the canonical commutation relation between the Interaction picture (IP) and Heisenberg picture (HP) if we work in the non-relativistic quantum mechanics (finite degrees of freedom). In this sense, each hermitian operator matches the same empirical result from cross section scattering. Nonetheless, the situation is not the case if switching the framework to quantum field theory (infinite degrees of freedom), furthermore, we cannot assure the existence of such an equivalent representation. In addition, the famous Haag's theorem \cite{Haag} undermines the perturbative evaluation of operators in the IP approach since there is no unitary representation from the Hilbert space of free and interacting vacuum state. To avoid this mathematical inconsistency, we choose to capture the perturbative scattering calculation under the Heisenberg representation.

The preceding declarations denote that a coherent alternative to solve quantum electrodynamics still misses. In a series of papers, Källén preferred to tackle $QED$ problems in four spacetime dimensions in the HP \cite{Kallen1,Kallen2}. Instead of working directly with Hamiltonian, the Källén method applies an expansion for every operator in the HP as a power series in the gauge coupling and substitutes into the equations of motion to deduce the rules of $S$-matrix. Such method has been applied in Thirring model \cite{Lunardi}, Scalar Quantum Electrodynamics (SQED) \cite{Lunardi2}, quantum correction at one loop level in $QED_3$ \cite{Tomazelli}.

Motivated by these considerations, the main goal of this paper is to evaluate the differential cross section of M{\o}ller scattering in the $GQED_3$ within the framework of Heisenberg representation. This paper is outlined as follows. In section II, we begin a brief revision of $GQED_3$ and exhibit the free solution and gauge field propagator. In section III, we review Dirac quantization. We devote the main aspects of $S$-matrix and the calculation at second-order approximation in section IV. The main part of this work we evaluate the M{\o}ller scattering in section V. Finally, we dedicate the section VI to concluding remarks.

 







\section{General discussion}
\label{sec:1}

In this section, we intend to review the basic results of $GQED_3$ and obtain the analytic gauge propagators. The starting point is the Podolsky lagrangian density function in a generalized $d$ space-time dimensions with a matter-current density $j^{\mu}(x)$ coupled to a gauge field $A_\mu (x)$

\begin{equation}\label{llkk}
\mathcal{L}_{GQED_d} = -\frac{1}{4 e^2}F^{\mu\nu}F_{\mu\nu} - \frac{a^2}{2 e^2} \partial^\mu F_{\mu\beta}\partial^\alpha F_{\alpha\beta} + j_\mu A^\mu
\end{equation}

Here, we assume $A_{\mu}(x)$ as continuous partial derivative up to fourth order in a $d$-dimensional volume element in Minkowski space-time \footnote{ We should adopt the Greek indices run from $1$ to $3$ and natural units $\hbar = c = 1$.}. The field-strength tensor defined as $F_{\mu\nu} = \partial_\mu A_\nu - \partial_\nu A_\mu$ and its two-form leads to Bianchi identity $dF = 0$. The Podolsky mass is $m_P = a^{-1}$ and we can recover the ordinary $QED_d$ if taking the limit $m_P \rightarrow \infty$.

To help visualize the behavior of $GQED_3$, one adopts the Lagrangian above to reveal the dimensional peculiarity by a simple redefinition of couplings where the effective nature of coupling may show the dimension signature. 
We should then examine the effective coupling of Podolsky mass. A systematic look in $(2+1)$ and $(3+1)$ dimensions conduct to $(e' m_P')^2 =  (e m_P )^2 / E^3$ and $(e' m_P' )^2 = (e m_P)^2 / E^2 $, respectively, where the energy scale is $E$. To understand these physical consequences, we observe gauge field has $[A] = M$, in any dimension, with $M$ mass unity and the dimensionless of coupling $e$ in $GQED_4$ shows the effective coupling $[m_P'] = M$ and   $e'$ are independent contributions, while in $GQED_3$ they are connected because of $[ e^2 ] = M $. 

Furthermore, according to already known result in $QED_3$, which photon are free in higher energy and strongly coupled in low energy, we prove the $m_P$ can archive faster than $QED_3$ the asymptotic ultraviolet and infrared photon states. Therefore, $GQED_3$ serves as a ground framework to IR system as mechanism of confinement as pair cooper and quarks into hadrons.


Returning now to the Lagrangian field in eq.  \eqref{llkk} where  $d = 3$, the principle of least action states  

\begin{equation}\label{elpod}
(1 - a^2 \Box ) \partial_\mu F^{\mu \nu} = 0,
\end{equation}

where $\Box \equiv \partial^\alpha \partial_\alpha$ \footnote{ Using $x_3 = i x_o = ic t $ and $ - ds^2 = dx^2_\alpha $.} . As it is clear from the perspective of any gauge theory, we must demand the absence of irrelevant degrees of freedom by a suitable choice of gauge. In order to face this problem, Podolsky \textit{et all.} based on covariant assumption of $ME$ imposed the well known Lorentz gauge $ \Omega_L [A] = \partial_\mu A^\mu$ \cite{Podolskyorigin}. Indeed, $GQED$ presents new constrains on the gauge field due to fact there are more physical variables than ME, furthermore, we can argue that eq. \eqref{elpod} and its initial condition destroy the $\Omega_L [A]$ constrain \cite{galvao}. Following the form of \eqref{llkk}, we would guess $\Omega_G [A]=(1 - a^2\Box)\partial_\mu A^\mu $ as a natural choice, nevertheless, the main evidence against shows the order of equations of motion are no longer preserved. Continuing in this direction, we accomplish the problem of gauge fixing by the notorious \textit{nonmixing gauge} $\Omega_P [A]=(\sqrt{ 1 - a^2 \Box }) \partial_\mu A^\mu $ \cite{nomixing1}. Therefore, incorporating the gauge framework $\Omega_P [A]$ into the \eqref{llkk} with $ d = 3 $, we have

\begin{equation}\label{llkk2}
\mathcal{L}_{GQED_3} = -\frac{1}{4}F^{\mu\nu}F_{\mu\nu} - \frac{a^2}{2} \partial^\mu F_{\mu\beta}\partial^\alpha F_{\alpha\beta} + \frac{1}{2}(\partial_\mu A^\mu)(1-a^2 \Box)(\partial_\mu A^\mu).
\end{equation}
 
The equations of motion now are
 
\begin{equation}\label{1532}
(1 - a^2 \Box ) \Box A^\mu (x) = 0.
\end{equation}

Here, the main aspect of free theory is that we can split  only in the free case the gauge field into two families of solutions $ A^\mu (x) = A_{Max}^\mu(x) + A_{Pod}^\mu (x) $, where $A_{Max}^\mu(x)$ and $A_{Pod}^\mu (x)$ are Maxwell and Podolsky gauge field, respectively. Within this factorization, each dynamical field is further understood upon equation \eqref{1532} as   

\begin{equation}\begin{aligned}\label{era}
(1 - a^2 \Box ) A^\mu (x) &= A_{Max}^\mu (x), \quad a^2 \Box A^\mu (x) =  A_{Pod}^\mu (x) \\
(1 - a^2 \Box ) A_{Pod}^\mu (x) &= 0, \quad \quad \quad   \quad
\Box A_{Max}^\mu (x) = 0 
\end{aligned}\end{equation}

One can immediately notice that there are modes with different dispersion relations and the physical interpretation of a massive and massless photon suits well to a non interacting photon. In addition, we introduce the general solution of \eqref{1532} in momentum space 

\begin{equation}\begin{aligned}
A_\mu (x) = \int \frac{d^2 \textbf{p}}{(2\pi)}   \sum^3_{\lambda=1}   \bigg\{ \epsilon_\mu^\lambda (p) ( a({\bf p}) e^{ipx} + a^* ({\bf p}) e^{- i p x}) + \eta_\mu (p) \bar{a}({\bf p}) e^{i \bar{p} x} + \eta_\mu^* (p) \bar{a}^* ({\bf p}) e^{- i \bar{p} x}) \bigg\}
\end{aligned}\end{equation}

where $p_\alpha = ({\bf p},i p_o)$ and $\bar{p}_\alpha = ({\bf p},i \bar{p}_o)$ with $ p_o = {\bf p}$ and $\bar{p}_o = (1 + a^2 p_o^2)^{1/2}/a $. The only nonvanishing commutation relations are $[a^\lambda ({\bf p}), \ a^{* \lambda'} ({\bf p}') ] = \delta_{{\bf p},{\bf p}'} \delta^\lambda_{\lambda'} = [\bar{a}^\lambda ({\bf p}), \ \bar{a}^{* \lambda'} ({\bf p}') ]$. The massless and massive polarization vector are $\epsilon_\mu^\lambda(p)$ and $\eta_\mu (p)$, respectively, with $\epsilon
^{\mu \lambda} (p) \epsilon_\mu^{\lambda'} (p) = \delta^{\lambda' \lambda}$ and $\eta^\mu (p) \eta_\mu^{*} (p) = - 1 $. Now, we demand the construction of free propagator in terms of the commutation relation of the gauge field $A_\mu (x)$ at equal time 

\begin{equation}\label{qwe2}
[A_\mu (x), A_\nu (x')] = -i \delta_{\mu\nu} D_P (x'-x)
\end{equation}

where the expression in momentum space is

\begin{equation}\label{derw}
D_P (x'-x) = \frac{-i}{(2\pi)^2} \int d^3 p e^{ip(x'-x)}(\delta^3(p^2) - \delta^3(p^2 + a^{-2}))\epsilon(p). 
\end{equation}


The propagator is a $c$-number. 
Subsequently, we are interesting in the Feynman propagator. Before doing so, it is important to separate \eqref{derw} into the retarded $D_P^R (p^2) = - \Theta(x_o) D_P(p^2)$  and advanced  $D_P^A (p^2) = \Theta(- x_o) D_P (p^2)$ propagator  

\begin{subequations}\begin{align}
\label{retb}
D^R_{P} (x'-x) &= \frac{1}{(2 \pi)^3} \int d^3 p e^{ip(x'-x)} \bigg( \mathcal{P} \frac{1}{p^2} - \mathcal{P} \frac{1}{p^2 + a^{-2}} + i \pi (\delta(p^2) - \delta(p^2 + a^{-2})) \epsilon(p) \bigg), \\
 \label{avcb}
D^A_{P} (x'-x) &= \frac{1}{(2 \pi)^3} \int d^3 p e^{ip(x'-x)} \bigg( \mathcal{P} \frac{1}{p^2} - \mathcal{P} \frac{1}{p^2 + a^{-2}} - i \pi (\delta(p^2) - \delta(p^2 + a^{-2})) \epsilon(p) \bigg),
\end{align}\end{subequations}

where $\mathcal{P}$ means the principal value, and $\epsilon(p)$ and $\Theta(x_o)$ are defined as
 
\begin{equation}
    \epsilon (p) \equiv \frac{p_o}{|p_o|}, \quad \Theta (p) \equiv \frac{ 1 + \epsilon (p)}{2}.
\end{equation}

For completeness, we should determine the vacuum expectation value of anticommutation relation at equal time 
 
\begin{equation}\label{qwe}
\langle 0 | \{ A_\mu (x), A_\nu (x')  \} | 0 \rangle = \delta_{\mu\nu} D^{(1)}(x'-x)
\end{equation}

where

\begin{equation}
D^{(1)}_P (x'-x) = \frac{1}{(2\pi)^2} \int d^3 p e^{ip(x'-x)} ( \delta^3 (p^2) - \delta^3 (p^2 + a^{-2}) )
\end{equation}

The equations \eqref{qwe2} and \eqref{qwe} can be further  combined into the Feynman propagator  

\begin{equation}\begin{aligned}\label{boring}
D^F_P (x) (x'-x) &= -\frac{1}{i}\epsilon(x'-x) D_P (x'-x) +  D_P^{(1)}(x'-x) = \frac{2}{i} \frac{1}{(2 \pi)^3} \int d^3 p e^{ip(x'-x)} \bigg\{ \mathcal{P} \frac{1}{p^2} \\ 
& - \mathcal{P}\frac{1}{p^2+a^{-2}} + i \pi ( \delta^3(p^2) - \delta^3(p^2 + a^{-2}) )\bigg\}.
\end{aligned}\end{equation}

In fact, this propagator regards as a superposition of positive and negative frequency propagating in the future and past light cone, respectively. As we should see later, this causal correlation function will be a fundamental ingredient, in section \ref{555}, to explore the M{\o }ller scattering. Moreover, we are in a position to solve an inhomogeneous differential equation by elementary methods \cite{scharf}. From the Lagrangian \eqref{llkk}, we find the Euler-Lagrange equation 

\begin{equation}\label{ihf2} 
(1 - a^{-2} \Box) \ \Box A^{\mu} (x) = j^\mu (x)
\end{equation}

where the general solution is 

\begin{equation}\label{examp}
A_{\mu} (x) = A_{\mu}^{(0)} (x) + \int d^3 x' D_R(x-x') j_\mu (x')     
\end{equation}

where $A_{\mu}^{(0)} (x)$ obeys the free equation \eqref{1532}. 
In the next section, we should address the basic structures of Fermion propagator in HP.


\section{The Dirac Field}\label{333}

To calculate the transition amplitude of M{\o}ller scattering, we should consider the Fermionic sector for a single spinor governed by the charge-symmetric Dirac Lagrangian 
 
\begin{equation}\label{Diracfree}
\mathcal{L}_{\psi} = - \frac{1}{4} [\bar{\psi}, \ ( \gamma \cdot \partial + m ) \psi   ] - \frac{1}{4} [\bar{\psi} ( \gamma \cdot \overleftarrow{\partial} + m ), \ \psi   ]
\end{equation}
 
where $\bar{\psi} \equiv \bar{\psi}_b (x)$ and $\psi \equiv \psi_b (x)$ are Grassmannian operators \footnote{The spinor indices $b$ runs from $1$ to $2$.}. By the Hamilton's principle, we obtain the dynamic of the system


\begin{subequations}\begin{align}\label{fdsw}
( \gamma_\mu^{ab}  \partial^\mu + \delta^{ab}  m ) \ \psi_b (x)=0, \\ \label{fdsw2}
\bar{\psi}_a (x) \  ( \gamma_\mu^{ab}  \overleftarrow{\partial}^\mu + \delta^{ab}  m ) =0
\end{align}\end{subequations}
 
being $\overline{\psi}_a (x)= \psi^*_a (x)\gamma^0 $ where $\gamma^\mu = ( i \sigma^3, \ \sigma^1, \  \sigma^2 )$ with Pauli matrix $\sigma^i$ \cite{scharf}. The solutions for equations of motion above are 

\begin{equation}\begin{aligned}\label{dirac123}
\psi_a (x) &= \sum^2_{r=1} \int\frac{d^3p}{(2\pi)} ( \hat{a}^r (\mathbf{p}) u_a^r(\mathbf{p})e^{-ipx} + \hat{b}^{\dag r}(\mathbf{p})  v_a^r(\mathbf{p})e^{ipx} ), \\ 
\overline{\psi}_a (x) &=\sum^2_{r=1} \int\frac{d^3p}{(2\pi)}( \hat{a}^{\dag r}(\mathbf{p})\bar{u}_a^{r}(\mathbf{p})e^{ipx} +  \hat{b}^r (\mathbf{p})\bar{v}_a^{r}(\mathbf{p})e^{-ipx} )
\end{aligned}\end{equation}

where $\bar{u}(\textbf{q})= u^* (\textbf{q}) \gamma^0 $. In order to maintain the stability and unitarity of the free system, the creation and annihilation operators of particle $(a^{r},a^{\dag r})$ and antiparticle $(b^r , b^{\dag r})$, respectively, must obey the Fermi-Dirac distribution from the spin-statistic theorem \cite{scharf}, and thus fulfill the commutation rules at equal times given by

\begin{equation}\label{comopcreani}\begin{cases}
\{ \hat{a}(\mathbf{p}),\hat{a}^\dag(\mathbf{p}') \} =\delta(\mathbf{p}-\mathbf{p}' ),\\
\{ \hat{b}(\mathbf{p}),\hat{b}^\dag(\mathbf{p}') \} =\delta(\mathbf{p}-\mathbf{p}' ).\\
\end{cases}\end{equation}

Moreover, we should formulate the necessary tools to interpret and calculate the perturbative element of $S$-matrix. We proceed in complete analogy with electromagnetism developed in the previous section. Hence, one can see explicitly the anticommutation of fermion operators at equal time through \eqref{comopcreani}, then 
  
\begin{equation}\label{cmd}
\left\{ \Bar{\psi}_a(x), \ \psi_b (x') \right\} = -i S_{b a } (x'- x). 
\end{equation}

Note that $S_{b a }$ is a invariant $c$-number and only a matrix in spinor space, where the expression in momentum space is

\begin{equation}\label{spinin}
S_{a b }(x-x')  = \frac{-i}{(2\pi)^3 } \int d^3p e^{ip(x-x')}  (i \gamma p - m)_{a b }   \delta^3 (p^2 + m^2) \epsilon(p)
\end{equation}

with a proper initial condition $S_{a b }$ is a free solution of Dirac equation $(\gamma \cdot \partial + m ) S_{a b } (x)= 0 $. In a similar fashion, we take the vacuum expectation value of spinor field commutation since one verifies the $[\Bar{\psi}_a (x) , \ \psi_b (x')]$ is no longer a $c$-number. 

\begin{equation}\label{s11}
\langle 0 | [\Bar{\psi}_a (x) , \ \psi_b (x')] | 0 \rangle  = S^{(1)}_{b a} (x'- x) 
\end{equation}
 
where the matrix $S^{(1)}_{b a}$ is 

\begin{equation}\begin{aligned} \label{s1}
S^{(1)}_{b a}(x - x')  &=  \frac{1}{(2\pi)^2} \int d^3p e^{ip(x-x')} (i \gamma p - m)_{b a}  \delta(p^2 + m^2).
\end{aligned}\end{equation}

Alternatively, we can identify the equation \eqref{cmd} as superposition of retarded $ S_R (x) = -\Theta(x_o) S (x)$ and advanced $ S_A (x)=\Theta(- x_o)S(x)$ propagators, for sake of simplicity the spinor indices are implicit. We then obtain the integral representation

\begin{subequations}\begin{align}
\label{retf}
S_R(x) &= \frac{1}{(2\pi)^3}
\int d^3p e^{ixp} (ip\gamma - m)[P \frac{1}{p^{2}+m^2} + i \pi \epsilon(p) \delta(p^{2} + m^2) ],  \\
\label{avcf}
S_A(x) &= \frac{1}{(2\pi)^3}
\int d^3p e^{ixp}(i p \gamma - m)[P \frac{1}{p^{2}+m^2} - i \pi \epsilon(p) \delta(p^{2} + m^2) ]. 
\end{align}\end{subequations}

It will be clear that these representations together with eqs. \eqref{retb} and \eqref{avcb} are the base to solve scattering amplitude in the framework of the Heisenberg picture. We are now in the position to briefly drawn the general structure from the interacting system. We begin with  

\begin{equation}\label{ihf}
(\gamma \cdot \partial + m)\psi(x) = g(x)
\end{equation}

where $g(x)$ is a product of field operators satisfying the requirement of locality and relativistic invariance.  It is possible to check with the aid of the standard techniques of the Green function and adequate boundary condition that the solution can be written as

\begin{equation}\label{12qw}
    \psi (x) = \psi^{(0)}(x) - \int d^3 x' S_R (x-x') g(x') 
\end{equation}
 
where $\psi^{(0)}(x)$ is the solution of \eqref{fdsw}. It is natural to explore the conjugated equation from \eqref{ihf}

\begin{equation}\label{wtf2}
\bar{\psi} (x)  (  \gamma \cdot \overleftarrow{\partial} +  m ) = g(x)
\end{equation}

where the general solution reads

\begin{equation}\label{13qw}
\bar{\psi} (x) = \bar{\psi}^{(0)}(x) - \int d^3 x'  g(x') S_A (x'-x) \end{equation}

where $\bar{\psi}_a (x)$ solves the eq. \eqref{fdsw2}. 
According to Noether's theorem in $QED$, the conservation of charge is connected with the abelian symmetry group $U(1)$. The properties of the field in $GQED$ left the abelian symmetry unchanged and requires by Lorentz invariant reason that the lagrangian of particles and interaction are equal to ME. As it is clear from $U(1)$, the status of symmetry naturally opens the way for the charge current density becomes $j^{\mu} = e \bar{\psi} \gamma^\mu \psi $ \cite{scharf} but we can also reduce the current to guarantee a symmetrized definition \cite{Kallen1,Kallen2}

\begin{equation}\label{CORRP}
j_{\mu} (x) \equiv \frac{ie}{2} \ [\Bar{\psi}(x), \gamma_\mu \psi(x)]
\end{equation}
 
this approach turns out to be convenient and should be instructive writing down explicitly the quantized operator current into normal products $ \frac{1}{2}[\overline{\psi} (x), \gamma_{\mu} \psi (x)] =  :\overline{\psi} (x) \ \gamma_{\mu} \ \psi (x): $ 
where "::" corresponding to chronological ordering \cite{scharf}. It is straightforward to notice the vacuum expectation value of this quantized current vanishes  

\begin{equation}\label{4corrienteq2} 
\langle 0|j_\mu(x)|0 \rangle = 0.
\end{equation} 
 
Before attempting to construct the M{\o}ller amplitude in the Podolsky frame, we should discuss the perturbative methods to find out the $S$-matrix.

\section{Heisenberg picture}\label{444}


The objective of this section is to describe the $S$-matrix, which relates the structure of the physical process, in the Heisenberg representation in order to circumvent the inconsistency already pointed out by Haag's theorem \cite{Haag}. Here, we follow the Källén methodology \cite{Kallen1,Kallen2} to shed some light on the cross section of $GQED_3$ experiments. One of the approaches towards a description of $S$-matrix is through the perturbative framework, in other words, expressing the matrix as a series in the power of small coupling. We therefore focus our attention on how to build up the perturbative terms of $S$-matrix in the HP for $GQED$, which has a straight extension from $QED$.
 
Before beginning, the idea behind the IP is that we may separate the total Hamiltonian into a free and an interacting part. On the other hand, the HP admits one analogous process where we can construct each operator in the Heisenberg representation as a homogeneous and inhomogeneous superposition of the solution equation \cite{Suer}. 

We now move on to present how $S$-matrix originates in the HP. To do this, we are interested in the differential equations of motion without any reference to the spacelike surface as in IP \cite{Suer,dysonn}. Summing over eqs. \eqref{llkk2}, \eqref{Diracfree}, and the minimal coupling $j^\mu (x) A_\mu (x)$, where $j^{\mu}(x)$ is \eqref{CORRP}, we arrive at dynamics

\begin{equation}\begin{aligned}\label{r122}
(1 - a^2 \Box ) \ \Box A_\mu (x)  &= - \frac{ie}{2} [\bar{\psi} (x) , \ \gamma_\mu \psi (x)] = - j_\mu (x), \\ 
(\gamma \cdot \partial + m) \ \psi(x) &= i e A_\mu (x) \gamma^\mu \psi (x).  
\end{aligned}\end{equation}


From systematic methods used in the previous sections, we can write down the set of solution for equations of motion above





\begin{equation}\begin{aligned}\label{sdfgh}
\psi (x) = \psi^{(in)}(x) - i e \int d^3 x' S_R (x-x') \gamma_\nu A^\nu (x') \psi (x') \\
A_\mu (x) = A_\mu^{(in)} (x) + \frac{i e}{2} \int d^3 x' D^R_P (x-x')[ \ \bar{\psi} (x') , \gamma_\mu \psi (x') \ ]   
\end{aligned}\end{equation}

Since the dynamical variable above contain retarded functions, we can associate the incoming field operators $( A_\mu^{(in)}, \ \psi^{(in)})$ as the initial values of $ ( A_\mu (x), \ \psi (x) )$ at $x_o \rightarrow - \infty$. Quite naturally, we possible consider solving the differential equations \eqref{r122} by advanced singular function

\begin{equation}\begin{aligned}\label{asdfg}
\psi (x) &= \psi^{(out)}(x) - i e \int d^3 x' S_A (x-x') \gamma_\nu A^\nu (x')  \psi(x') \\
A_\mu (x) &= A_\mu^{(out)} (x) + \frac{ie}{2} \int d^3 x' D^A_P (x-x') [\bar{\psi} (x'), \gamma_\mu \psi (x')]
\end{aligned}\end{equation}

where $( A_\mu^{(out)}, \ \psi^{(out)})$ are defined as the limit $x_o \rightarrow + \infty $ for free field operator $( A_\mu^{(0)} \ , \ $ $  \psi^{(0)})$ and often called "outgoing" fields. Physically, they corresponds to the final value of $(A_\mu \ , \ \psi )$ when we switching off adiabatically the interaction. In other words, the interaction vanishes in the limit $x_o \rightarrow + \infty $ and so the  operator are governed by free field equations. 

In addition, it is also essential that incoming and outgoing fields obey the same commutation relation of free-fields

\begin{minipage}[t]{0.5\textwidth}
\begin{equation*}\label{aqws2}\begin{cases}
\{ \bar{\psi}_a^{(in)} (x), \ \psi_b^{(in)} (x') \} \ = \ - i S_{ba}(x'-x),\\
[A_\mu^{(in)}(x), \ A_\nu^{(in)}(x')] = -i \delta_{\mu\nu} D_P (x'-x),\\
[A_\mu^{(in)}(x), \ \psi_b^{(in)} (x')] = 0,
\end{cases}\end{equation*}
\end{minipage}
\begin{minipage}[t]{0.5\textwidth}
\begin{equation*}\label{aqws}\begin{cases}
\{ \bar{\psi}_a^{(out)} (x), \ \psi_b^{(out)} (x') \} \ = \ - i S_{ba}(x'-x),\\
[A_\mu^{(out)}(x), \ A_\nu^{(out)}(x')] = -i \delta_{\mu\nu} D_P (x'-x),\\
[A_\mu^{(out)}(x), \ \psi_b^{(out)} (x')] = 0.
\end{cases}\end{equation*}
\end{minipage} 

\vspace{4mm}

The justification of these commutation rules regards the fact that the free, incoming, and outgoing operators share the differential equations of motion \eqref{r122} without interactions. This scenario also illustrate the free field operators converge at $ ( x_o \rightarrow - \infty )$ and $ ( x_o \rightarrow + \infty )$ to incoming and outgoing fields, respectively, \cite{Suer}.



We thus reach, without difficult, an important conclusion where a canonical transformation must connect the asymptotic fields $(A_\mu^{(in)} \ , \ \psi^{(in)})$ and $(A_\mu^{(out)} \ , \ \psi^{(out)})$ since both of them settle down the same canonical commutation relation. Then it follows

\begin{equation}\begin{aligned}\label{sus}
\psi^{(out)} (x) &= S^{-1} \ \psi^{(in)} (x) \ S  \\
A_\mu^{(out)} (x) &= S^{-1} \ A_\mu^{(in)} (x)  \ S
\end{aligned}\end{equation}

the operator $S$ must be unitary since it relates two sets of orthogonal operators  

\begin{equation}\label{um}
S S^* = S^* S = \mathbb{1}    
\end{equation}

In analogous fashion with \eqref{sus}, we emphasize the Hamiltonian at $x_o \rightarrow + \infty$ can be decomposed as function of the Hamiltonian at $x_o \rightarrow - \infty$ in the form

\begin{equation}
H^{(0)} (\psi^{(out)},A_\mu^{(out)}) = S^{-1}  H^{(0)} (\psi^{(in)},A_\mu^{(in)})  S,  
\end{equation}

then one should be aware that the incoming and outgoing fields live in the same Hilbert space \cite{Kallen1,Kallen2}. 
At this point, we establish a general formulation as a first step towards the perturbative formulation of the scattering matrix by replacing eqs. \eqref{sdfgh} into \eqref{asdfg} to get

\begin{equation}\begin{aligned}\label{sssss}
& \psi^{(out) }(x) \ = \ S^{(-1)} \ \psi^{(0)} (x) \ S = \ \psi^{(0)}(x) - i e \int d^3 x' S (x-x') \gamma_\nu A^\nu (x') \psi (x') \\
& A_\mu^{(out)} (x) \ = \ S^{(-1)} \ A_\mu^{(0)} (x) \ S = \ A_\mu^{(0)} (x) + \frac{i e}{2} \int d^3 x' D_P (x-x')[\bar{\psi} (x') , \gamma_\mu \psi (x')] 
\end{aligned}\end{equation}

or rewriting to have a systematic view of $S$-matrix as

\begin{equation}\begin{aligned}\label{ss}
[S, \ \psi^{(0)}] = \ - S \int d^3 x' S(x-x') i e \gamma^\mu A_\mu (x') \psi (x'), \\
[S, \ A_\mu^{(0)}] = \ - S \int d^3 x' D(x-x') \frac{i e}{2} [ \bar{\psi} (x') , \gamma_\mu \psi(x') ]
\end{aligned}\end{equation}

For practical computation and without losing the physical meaning, we adopted the notation $( A^{(0)} \ , \ \psi^{(0)} )$ to incoming fields. Using the assumption of the small gauge coupling $( e^2/4 \pi \sim 1/137 )$ and expanding the operators $( S, \ \psi, \ A^\mu) $ in power series of $e$. We get


\begin{equation}\begin{aligned}\label{P1}
S &= \mathbb{1} \ + \ e S^{(1)} + \ e^2 S^{(2)} \ + \ \dots \\
\psi(x)&=\psi^{(0)}(x)+ \ e\psi^{(1)}(x) \ + \ e^{2}\psi^{(2)}(x) \ + \ldots,\\
A_{\mu}(x)&=A_{\mu }^{(0)}(x)+ \ eA_{\mu }^{(1)}(x) \ + \ e^{2}A_{\mu }^{(2)}(x)+ \ldots .
\end{aligned}\end{equation}

\vspace{2mm}

Substituting \eqref{P1} into \eqref{ss}, the first approximation reads

\begin{subequations}\begin{align}
\label{Fist}
[S^{(1)}, \ \psi^{(0)}(x)] \ = \ - S \int d^3 x' S(x-x') \ i e \gamma^\mu A^{(0)}_\mu (x') \psi^{(0)} (x'), \\
\label{sb}
[S^{(1)}, \ A_\mu^{(0)}(x)] = \ - S \int d^3 x' D(x-x') \ \frac{i e}{2} [ \bar{\psi}^{(0)} (x') , \gamma_\mu \psi^{(0)} (x') ].
\end{align}\end{subequations}

After some manipulation, the form of $S$-matrix is

\begin{equation}
S^{(1)} = \ - i e \int d^3 x :\bar{\psi}^{(0)}(x) \gamma^\mu \psi^{(0)}(x):A^{(0)}_\mu (x).    
\end{equation}

To help visualize this result, see eq \eqref{qwe2}. The calculation for the second order of $S$-matrix is straightforward

\begin{equation}\begin{aligned}\label{qwsa}
S^{(2)} &= \ \frac{e^2}{4} \int d^3 x' d^3 x'' \ T ( : \bar{\psi}^{(0)}(x') \gamma^{\nu_1} \psi^{(0)}(x') : : \bar{\psi}^{(0)}(x'') \gamma^{\nu_2} \psi^{(0)}(x'') : ) \\ 
& \times  T( A^{(0)}_{\nu_1} (x') A^{(0)}_{\nu_2} (x'')).
\end{aligned}\end{equation}

Despite the considerations above appear complicated, the $S^{(n)}$ terms are very direct. 
We should underline that the link between HP and IP has only the same mathematical structural form of perturbative expansion but the underlying physical concepts involved turn these framework into different address scattering matrix. The advantage of the HP is clear, there is no need to recover the space-like surfaces \cite{dysonn}.  

In the remainder of this paper, the core idea from Haag's theorem is that the free and interacting Hamiltonian from IP act on orthogonal Hilbert spaces, in other words, the mathematical inconsistency resumes in the lack of a global unitary transformation relating both of them \cite{Haag}, as the argument shown in eq. \eqref{sus}. Thus, the $S$-matrix in the HP is not ruined by Haag's theorem and thus the HP framework is appropriated to carry out scattering process.

In the next section, we are ready to calculate the analytical solution of M{\o}ller scattering at tree-level.  


\section{M{\o}ller scattering}\label{555}

Now that we are more familiar with the Podolsky theory and display essential ideas to deal with the scattering matrix. We should move on to a better understanding of the scattering process in $GQED_3$. We focus on showing how Källén methodology \cite{Kallen1,Kallen2} in the HP is an attractive viewpoint to address the perturbative apparatus of the scattering process without worrying about Haag's theorem. The main objective is to determine the Podolsky corrections for M{\o}ller scattering $(e^- e^- \rightarrow e^- e^-)$ at tree-level. Starting with the incoming $| \ p , \ q \rangle $ and outgoing $| \ p', \ q' \rangle$ states of electrons

\begin{equation}\begin{aligned}
| \ p , \ q \rangle &= a^{*(r_i)}(p) \ a^{*(s_i)}(q) | 0 \rangle \\
| \ p', \ q' \rangle &= a^{*(r_f)}(p') \ a^{*(s_f)}(q') | 0 \rangle
\end{aligned}\end{equation}

where $(p,r_i)$ and $(q,s_i)$ are the momentum and spin of ingoing particle, and $(p',r_f)$ and $(q',s_f)$ are the momentum and spin of outgoing particle. Being $| 0 \rangle$ the vacuum state. By the spin-statistics theorem, the interchanges of identical particles must follows the rule 

\begin{equation}\label{fdg}
| p, \ q \rangle = -  |q, \ p \rangle .
\end{equation}

After that we are able evaluate the cross section for the process $e^- \ e^- \rightarrow e^- \ e^-$. The first non-vanishing approximation involves the second order of $S$-matrix element \eqref{qwsa}, mapping the asymptotic initial states onto final ones 

\begin{equation}\begin{aligned}\label{dasdas}
\langle q', p'| S | p , q \rangle &= - \frac{e^2}{2}
\int d^3 x' d^3 x'' [ \  \langle q' |  \bar{\psi}^{0}(x') | 0 \rangle \gamma^{\nu_1} \langle 0 | \psi^{0} (x') | q \rangle \langle p' | \bar{\psi}^{0}(x'') | 0 \rangle \gamma^{\nu_2} \langle 0 | \psi^{0}(x'') | p \rangle \\ 
& - \langle q' |  \bar{\psi}^{0} (x') | 0 \rangle  \gamma^{\nu_1} \langle 0 | \psi^{0} (x') | p \rangle \langle p' |  \bar{\psi}^{0}(x'') | 0 \rangle \gamma^{\nu_2} \langle 0 | \psi^{0}(x'') | q \rangle  \  ] \delta_{\nu_1 \nu_2} \ D^F_P (x'-x''). \\
\end{aligned}\end{equation}

Substituting the Feynman propagator \eqref{boring} and matrix elements \eqref{dirac123}, we obtain 

\begin{equation}\begin{aligned}\label{moll}
\langle q', p'| S | p , q \rangle &= \frac{ i e^2 }{A^2} \ \bigg[  \frac{\bar{u} (q') \gamma_\lambda u (q) \bar{u} (p') \gamma_\lambda u (p)}{(p-p')^2 (1 + \frac{(p-p')^2}{m^2_P} )} \ - \ \frac{\bar{u} (q') \gamma_\lambda u (q) \bar{u} (p') \gamma_\lambda u (p)}{(p-q')^2 (1 + \frac{(p-q')^2}{m^2_P} )} \bigg] \ \times \\
& (2 \pi)^3  \delta^3 ( p + q - p' - q' )   
\end{aligned}\end{equation}

where $A$ is the area. In what follows, the omission of factor 2 occurs because of two possibilities for the combination of operators $\bar{\psi}(x')$ and $\bar{\psi}(x'')$ and the negative signal proceeds from eq. \eqref{fdg}. In the present context, the Podolsky propagator above shows the separation in two different families of the solution is no longer possible for interacting system, as we discussed in the eq. \eqref{era} .

We can infer the leading order of $S$-matrix elements in HP may be identified with the Feynman rules and thus the interpretation suggests a possible one-to-one interplay between eq. \eqref{dasdas} and Feynman diagrams
. Nevertheless, as we well know, the standard quantum field theory in IP relies on the direct association between Feynman techniques and expansion "a lá" Dyson. This statement has not been proved mathematically and remains more ignored when infrared divergences appear in the Feynman amplitude. We therefore continue to work with the position specified in the Heisenberg representation to obtain the differential cross section.



Now, we should infer the construction of the cross section with sufficient care to conduct the situation from a reduced dimension standpoint. The operation to determine this physical variable of interest could be applied in the same way as $(3+1)$ dimension. After taking the probability per unit of time from \eqref{moll} and dividing by the flux of incident particle $v_{rel}/A $ (flux rate across a curve), where $v_{rel}$ is the relative velocity. The incoming and outgoing states are then normalized to one particle per unit of area. We rather get

\begin{equation}\begin{aligned}\label{cs}
\sigma &= \frac{\alpha^2}{v_{rel}} \int \int \frac{d^2 p' d^2 q'}{16 E_p E_q E_{p'} E_{q'}} \bigg[\frac{A}{(p-p')^4 (1 + \frac{(p-p')^2}{m^2_P})^2 } +  \frac{B}{(p-q')^4 (1 + \frac{(p-q')^2}{m^2_P})^2} + \\
& \frac{C}{ (p-p')^2 (p-q')^2 (1 + \frac{(p-p')^2}{m^2_P}) (1 + \frac{(p-q')^2}{m^2_P}) }  \bigg] \delta^3 (p+ q - p' - q').
\end{aligned}\end{equation}

For sake of clarity, we averaged over ingoing spin states since they are unpolarized and sum of all possible spin state of outgoing particles. The $\alpha = e^2/4\pi$ is the fine constant structure. With the aid of the equations \eqref{dirac123}, the completeness relation of quantized wave equation are $\sum^2_{r=1} \bar{u}_\alpha u_\beta = (i \gamma \cdot q^+ - m)_{\beta \alpha}/2E $ and $\sum^2_{r=1} \bar{v}_\alpha v_\beta = -(i \gamma \cdot q^- - m)_{\beta \alpha}/2E$, where $q^+ = (\textbf{q}, iE)$ and $q^- = (\textbf{q}, -iE)$. The capital letters above indicate the summing over spin indices 

\begin{subequations}
\begin{align}
A & = Tr \left[ \gamma_{ \lambda }\left( 
i \gamma \cdot p - m \right) \gamma_{\nu }\left( i \gamma  \cdot p'  - m \right) \right] \cdot Tr \left[ i \gamma^{ \lambda } \left(
i \gamma \cdot q - m \right) \gamma^{\nu } \left( i \gamma \cdot q' - m \right) \right] ,\label{AAA} \\ 
B & =  Tr \left[ \gamma_{ \lambda }\left( 
i \gamma \cdot p - m \right) \gamma_{\nu }\left( i \gamma  \cdot q'  - m \right) \right] \cdot Tr \left[ i \gamma^{ \lambda } \left(
i \gamma \cdot q - m \right) \gamma^{\nu } \left( i \gamma \cdot p' - m \right) \right] ,\label{BBB} \\
C & = - 2 Tr \left[ \gamma_{ \lambda }\left( 
i \gamma \cdot p - m \right) \gamma_{\nu }\left( i \gamma  \cdot q'  - m \right) \gamma^\lambda  \left( i \gamma \cdot q - m \right) \gamma^{\nu } \left( i \gamma \cdot p' - m \right) \right]. \label{CCC}
\end{align}%
\end{subequations}

\vspace{2mm}

The new features of $GQED_3$ manifest in the properties of $\gamma$-matrices, in the sense, they obey the $so(1,2)$ algebra $[\gamma_\mu , \gamma_\nu] = - 2 i \epsilon_{\mu \nu \lambda} \gamma^{\lambda} $ and $\{ \gamma_\mu , \gamma_\nu \} = 2 g_{\mu\nu}$ \cite{scharf}. We now define the Mandelstam variables to describe future Lorentz invariant quantities   

\begin{subequations}
\begin{align}
s& = ( p+q ) ^{2}=( p'+q')^{2} = - 2m^{2} + 2( p.q) ,\label{3.7a}\\
t& =( p-p') ^{2}=( q-q')
^{2} = - 2m^{2} - 2( p.p') ,\label{3.7b}\\
u& =( p'-q) ^{2}=( p-q')^{2} = - 2m^{2}-2( p'.q) . \label{3.7c}
\end{align}%
\end{subequations}

\vspace{2mm}

Finally, after applying \cref{3.7a,3.7b,3.7c} into \cref{AAA,BBB,CCC} together with the algebra of Dirac matrices, we get

\begin{subequations}
\begin{align} 
& A = 2 (s^2 + u^2 - t^2/2) + 16 m^2 ( s + u - t/2) + 48 m^4 , \\
& B = 2 (s^2 + t^2 - u^2/2) + 16 m^2 ( s + t - u/2) + 48 m^4 , \\
& C = (5 s^2 - u^2 - t^2) + 8 m^2 (5 s - u - t) + 36 m^4 .
\end{align}
\end{subequations}

Next, one has to calculate the relative velocity to compute the scattering. The ordinary physical meaning of this concept infers as the velocity of a particle in relation to an observer in the rest frame of the other one. If the observer is at rest frame of $q$ particle. We have

\begin{equation}
v_{rel} E_p E_q = m E_p \frac{|\textbf{p}|}{E_p} = \sqrt{m^2 E^2_p - m^4 }.
\end{equation}

However, this argument is not physically well defined to relativistic phenomena. Adopting $v_{rel} E_p E_q$ in a relativistic invariant way. It is easy to see that

\begin{equation}\label{ver}
v_{rel} = \frac{1}{E_q E_p} \sqrt{(p \cdot q)^2 - m^4}.
\end{equation}




We now turn our attention to the analysis of the invariant differential cross section from eq. \eqref{cs} by using the "normalized" Lorentz invariant quantity $ \lambda_1 = t / 2m^2$ 
to get

\begin{equation}\label{ffggd}
\frac{d \sigma}{d t} = \frac{r^2_o}{32 \sqrt{ (p \cdot q)^2 - m^4}} \bigg[ \frac{A}{t^2 (1 + \frac{t}{m^2_P} )^2 } + \frac{B}{u^2 (1 + \frac{u}{m^2_P} )^2} + \frac{C}{ u t (1 + \frac{t}{m^2_P} ) (1 + \frac{u}{m^2_P} )} \bigg] \cdot I
\end{equation}


where the classical electron radius $r_o = \frac{\alpha}{m}$ is dimensionless. Let us first start with the Lorentz invariant phase space integral $I$ that is given by

\begin{equation}\begin{aligned}
I &= \int \int \frac{d^2 p' d^2 q' }{E_{p'} E_{q'}} \delta(p + q - p' - q') \delta^3 ( \lambda_1 - \frac{ t }{2 m^2}) \\
&= 4 \int \int d^3 p' d^3 q' \delta^3 (p'^2 + m^2) \delta^3 (q'^2 + m^2) \Theta (p') \Theta (q') \delta (p+q - p' - q') \delta^3 ( \lambda_1 - \frac{t}{2 m^2}) \\
&= \frac{2 m^2 }{\sqrt{4 t m^2 |\textbf{q}|^2 + s t^2}} \ \Theta (\frac{t}{2m^2}) \ \Theta(\frac{E_q}{m} - 1 - \frac{t}{2 m^2})
\end{aligned}\end{equation}

in the rest frame of the particle $p$. Moving to an arbitrary system of coordinate  

\begin{equation}
I = \frac{2m^2}{\sqrt{-s}\sqrt{tu}} \ \Theta(\frac{t}{2m^2}) \  \Theta(\frac{u}{2m^2}).    
\end{equation}

This result enables us to write the differential cross section \eqref{ffggd} in a relativistic invariant fashion. The influence of factor $(\sqrt{tu})$ concerns the transverse asymmetry even if the radiative correction from soft and hard photons are absent. This characteristic is reminiscent of $(2+1)$ dimensions, whereas in $(3+1)$ dimensions the effect is only supposed to be presented when infrared radiations are included \cite{Shumeiko}. Moreover, we recover the usual $QED_3$ cross section from \eqref{ffggd} in the limit $m_P \rightarrow \infty$ because the Podolsky Feynman propagator turns into the Maxwell one. The effects in the higher energy regime

\begin{equation}
m^{2} \ll  \left\vert s \right\vert  \sim  t \sim  u .
\end{equation}

Note that in this present situation, the form of Lorentz-invariant cross section \eqref{ffggd} is

\begin{equation}\label{vvbbn}
\frac{d \sigma}{d t} = \frac{\alpha^2}{(2\sqrt{-s})^3}\frac{1}{\sqrt{tu}} \bigg[ \frac{s^2 + u^2 - t^2/2}{t^2 (1 + \frac{t}{m^2_P} )^2 } + \frac{s^2 + t^2 - u^2/2}{u^2 (1 + \frac{u}{m^2_P} )^2} + \frac{5s^2 - u^2 - t^2}{ u t (1 + \frac{t}{m^2_P} ) (1 + \frac{u}{m^2_P} )} \bigg]. 
\end{equation}

The reader may wonder about this energy level where we find the direct influence of the Podolsky parameter. This regime is not interesting since the particles may archive high values of energy and the Podolsky mass cannot play the UV cutoff role. On the order hand, it leads immediately to a subtle investigation of the bound value of $m_P$. The next step will be favorable for our purposes since the particles have the energy value restrict to the following bounds

%







\begin{equation}
m^{2} \leq \left\vert s\right\vert \sim   t   \sim   u   <m_{P}^{2}.
\end{equation}

\vspace{1mm}

This regime points $m_P$ as a natural higher-frequency cutoff. Although this condition of energy could appear inadequate to conduct analysis in physical system of interest, we should recall the research in condensed matter with relativistic behavior is recent, for instance, considering the investigation on 
electron-electron interaction in graphene structure \cite{madrid}.

 
As a result of the leading contribution $\frac{\sqrt{t}}{m_P}$ in the eq. \eqref{vvbbn}, we have

\vspace{1mm}

\begin{equation}\begin{aligned}
& \frac{d \sigma}{d t} = \frac{\alpha^2}{(2\sqrt{-s})^3}\frac{1}{\sqrt{tu}} \bigg[ \frac{s^2 + u^2 - t^2/2}{t^2} + \frac{s^2 + t^2 - u^2/2}{u^2} + \frac{5s^2 - u^2 - t^2}{ut}  \\ 
& - \frac{3}{m_P^2} \bigg( \frac{3 s^2 + u^2 - t^2}{t} + \frac{3 s^2 + t^2 - u^2}{u} \bigg) \bigg].
\end{aligned}\end{equation}

\vspace{2mm}

It is quite interesting to evaluate in the center mass system

\begin{equation}
p = (E_i, \textbf{p}, \ 0), \ q =(E_i,  - \textbf{q} ,  0), \ p' = (E_f , \  {\bf \ p}  \cos \theta , \  {\bf p} \sin \theta \  ), \ q' = (E_f , \  - {\bf q} \cos \theta , \  - {\bf q} \sin \theta  )
\end{equation}

where $\theta$ is the scattering angle in the center of mass frame and energy-momentum conservation results $E_i = E_f$ and $ {\bf p} = {\bf q}$, i.e. $(\textbf{p} \cdot \textbf{q}) = p^2 \cos \theta$. Hence, we can rewrite the Mandelstam variable \eqref{3.7a}-\eqref{3.7c} as

\begin{equation}
s= - 4E^{2},\quad u=4\mathbf{p}^{2}\cos ^{2}\frac{%
\theta }{2},\quad t=4\mathbf{p}^{2}\sin ^{2}\frac{\theta }{2},  \label{3.10}
\end{equation}

we thus obtain the cross section in the center of mass

\begin{equation}\label{zzxxccvv}
\frac{d \sigma}{d \theta} = \frac{\alpha^2}{32 E^3} \frac{(7 + \cos 2 \theta)^2}{\sin^4 \theta} - \frac{\alpha^2}{m_P^2} \frac{3}{8 E } \frac{(7 - 6 \cos 2 \theta + \cos^2 2 \theta )}{  \sin^4 \theta }.
\end{equation}

At this point, the second term plays the central role of the correction in the relativistic limit from $GQED_3$, while the first one concerns the usual $QED_3$ contribution. Moreover, the small deviation for M{\o}ller scattering at tree-level cross section may be calculated with the formula

\begin{equation}\label{bhg}
\delta  = \left( \frac{d\sigma }{d\theta }\right)^{GQED_3}  \left/ \left( \frac{d\sigma }{d\theta }\right)\right. ^{QED_3} - 1.
\end{equation}
 
The most experimental advances in particle physics require the study of M{\o}ller scattering with higher precision even in higher energy reactions 
\cite{Aleksejevs}. The quantum ideas developed so far fit nicely with quasiplanar structures of condensed matter and renew the view of charge confinement and screening coulomb potential since a notorious quality of $GQED_3$ comes from the contribution of "Yukawa" positive electrostatic potential in $(-e/r)(1 - \textit{e}^{-r \ m_P})$ \cite{Cuzinattoww}. Then under certain circumstances we can achieve an attractiveness global interaction and expect the theoretical features of $GQED_3$ be significant to tackle with an effective description of condensed matter.

The leading order correction for small angles $\theta \ll 1$ in eq. \eqref{bhg}, we obtain 

\begin{equation}
\delta =-\left( \frac{s}{ m^2_{P} }\right) \frac{3}{4}\theta ^{2} + \mathcal{O} ( \theta^3 ) + \dots.
\end{equation}

The small angles report the kinematical region of low energy where the dominant process is $t$-channel. The $GQED_3$ lower order effect for the cross section in terms of small $\theta$ angle is 

\begin{equation}
\frac{d \sigma}{d \theta} = \frac{\alpha^2}{8 E^3 \theta^4} \bigg[ 1 - \bigg( \frac{1}{2} + \frac{E^2}{m_a^2} \bigg) \theta^2 + \mathcal{O} ( \theta^3 ) + \dots \bigg]
\end{equation}

In the above result, the $GQED_3$ contribution turns out at second order and decreases the differential cross section. 
We start by taking the electron mass $m=0,510~MeV$ and considering the energy limit where M{\o}ller scattering is relevant $< 100 \ MeV $   \cite{Epstein}. Thus, this constraint could inform the real accuracy of correction suggested by the generalized electrodynamics for $\theta = 10^\circ $, i.e., $|\delta| \leq 2. 10^{-6} \% $. An recent measurement of $e^- e^-$ in cold atoms lead us to $E \simeq 1.530 \ MeV$, within the estimation of M{\o}ller regime, and the order of correction for $\theta = 10^\circ $,  $|\delta| \leq 10^{-9} \%  $ \cite{coldd}. One could argue the experiments to M{\o}ller scattering are not appropriate to detect the Podolsky mass effects, otherwise it would already be detected.    

To discuss this challenge, we should map the phenomenological objects in different physical contexts by suitable choices. Since adopting $GQED_3$ as the analogous model for planar structure in the condensed matter system, we should link up the free parameters of theory with important observable. To carry this idea, we should therefore be able to identify the theoretical similarity between $GQED_3$ cutoff with experimental data. These results must be compatible with the standard interpretation of the planar quantum field theory.

Before attempting further progress, we remark the boundness limit of Podolsky parameter in $GQED_4$ was found from the anomalous magnetic momentum $m_P \geq 37.59$ GeV \cite{RBufBMP} and the ground state of Hydrogen $ m_P \geq 35.51 \ MeV$   \cite{Cuzinattoww}. To propose a correspondent model for condensed matter, we should replace the cutoff $m_P$ by the uncertainty in determining the interatomic distance since the known $ME$ results are in agreement with experiments that depend on interatomic distance in crystal lattice  \cite{citing}. So considering the measuring from \cite{Razik} we roughly argue $a \leq 3 fm $ or $m_P \geq 65.77 \ MeV$. It is therefore possible to  estimate the $GQED_3$ corrections, $|\delta| \leq 0.0001 \%$ to cold atoms for $\theta = 10^\circ$ \cite{coldd}.  
 


Finally, we explore the nonrelativistic limit where the long-distance physics is independent of $m_P$. In this approximation, the energy of particles are $E_q = E_p = E_{q'} = E_{p'} = m + \frac{p^2}{2m}$ and the cross section in the center of mass system reads

\begin{equation}\label{nonrela1}
\frac{d \sigma}{d \theta} = \frac{r_o^2}{16} \frac{m^3}{p^4} \bigg[ \bigg( \frac{4}{\cos^4 \frac{\theta}{2}} + \frac{4}{\sin^4 \frac{\theta}{2}} - \frac{11}{\cos^2 \frac{\theta}{2} \sin^2 \frac{\theta}{2}} \bigg) +  \frac{3 p^2}{m^2_P}  \bigg(  \frac{1}{\cos^2 \frac{\theta}{2}} + \frac{1}{\sin^2 \frac{\theta}{2}} \bigg)  \Bigg]   
\end{equation}

\begin{figure}[h]
 \centering
\includegraphics[height=0.30\textheight]{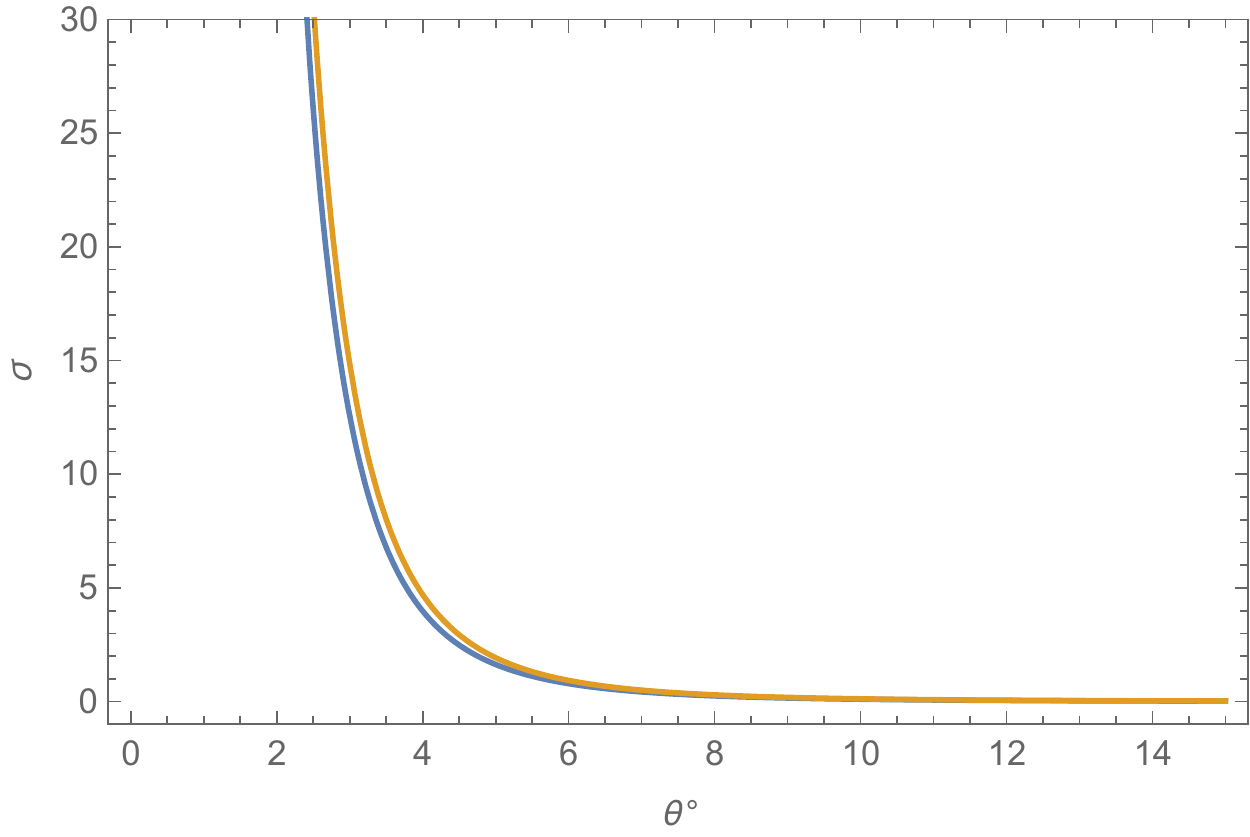}
\caption{\label{action} Scattering electrons compared with scattering angle. The blue and orange lines are the cross section using $\alpha = 2.3 $ and $\alpha = 2.5 $, respectively.}
  \label{fig:birds2}
\end{figure}

The first term represents the nonrelativistic ordinary $QED_3$, whereas the second one shows the leading "Yukawa" correction for the differential cross section from the massive contribution of Podolsky's potential  \cite{Podolskyorigin}. Judging the form of differential cross section, unlike what happens in the fermion sector where the algebra of gamma matrices set up the planar dynamic, the gauge propagator merely plays the same similar characteristic in $GQED_4$, namely, living in a spatial bulk space \cite{notre}.



We show a simulation for \eqref{nonrela1} in the graph of Fig. \ref{fig:birds2} based on the data \cite{madrid}, where the effective fine structure is $\alpha' = \frac{e^2}{4 \pi v_F} \sim 2.3 -2.5$ with the Fermi velocity $v_F$. The typical form relates a coulomb differential cross section in which diverges at forwarding angles.




It is essential to realize which aspects from a Chern-Simons (CS) term would imply in our model. 
We would like to emphasize which in $QED_3$ even though the CS is absent in the lagrangian \eqref{llkk2}, this term can be induced by photon self-energy at one loop order \cite{qsaco2}. Beyond that, we can find, without any loss of generality, the solution from the spontaneous breaking of parity, which is the same physical predictions if the CS is present in the eq.  \eqref{llkk2} \cite{qsaco}. Given the last remarks, if we take into account radiative correction for evaluating the differential cross section, we should expect a favorable scenario involving results from the induced 
CS term.

  



\section{Concluding Remarks}\label{666}

In this paper, we have analyzed the M{\o}ller scattering in the framework of generalized quantum electrodynamics in $(2+1)$ dimensions. One immediate feature of the $GQED$ framework was the unitarity and stability. We also derived the gauge propagator, which preserves the $U(1)$ symmetry, with a massless and massive photon . 

We have presented the covariant formalism of Källén's method where the $S$-matrix can be constructed directly in the Heisenberg picture. The main advantage of this representation was to remove the mathematical inconsistency barrier, so far demonstrated by Haag's theorem, associated with the canonical approach of the Interaction picture. We analyzed the structure of $S$-matrix, which was facilitated through the integral equation of motion, and collected the $S^{(2)}$ matrix amplitude to evaluate the tree-level M{\o}ller scattering.

The richness outcomes offered by $QED_3$ allowed an understanding of the elusive effects of condensed matter. However, due to a lack of concrete picture in some of these crucial phenomena, we proposed the $GQED_3$ framework where "short-ranged force" encoded by massive photon arisen naturally.  Moreover, even though $GQED_3$ easily copes with higher energy mechanism, Podolsky's theory was first investigated in the classical regime and the parameter $m_P$ played an important role in regularizing classical problems, as mentioned in the introduction. In this spirit, we hoped the cutoff parameter $m_P$ translated into condensed matter language could facilitate the investigation into the role of planar matter interactions.

The remarkable contribution in this paper was to show the M{\o}ller differential cross section in $QED_3$ and ensuing $GQED_3$. As we known, a topological mass is generated in quantum radiation at one loop order and if this mass exceeds the electron one, we have a favorable scenario to attractive potential (cooper pair) in condensed matter. Hence, we showed $GQED_3$ instead of $QED_3$ creates the proper conditions to work in the perturbative approach due to the energy regime: $m^2 \leq p^2 < m_P^2 $, where the cutoff is $m_P \geq 65.77 MeV $, and offer a better physical estimation of the future deviations from ME.

By fundamental arguments of quantum field theory, we paved the road to establish a systematic investigation of radiative effects in the differential cross section. We can extend this formalism to incorporate quantum fluctuations since we already known the principal ideas of generalized electrodynamics in the Heisenberg representation at tree-level process. We believe the higher orders of cross section from $GQED_3$ can also be interesting to understand the attractive mechanism of Cooper pair, charge confinement, dynamical mass generation and the connection of the Chern-Simons term with physical observable in the condensed matter. A further study in this formalism will be elaborated in a forthcoming work.




\subsection*{Acknowledgments}

David Montenegro thanks to CAPES for full support.


\end{document}